\documentstyle[12pt]{article}

\begin{document}

\setcounter{page}{1}

\pagestyle{plain} \vspace{1cm}
\begin{center}
\Large{\bf Non-Minimal Inflation on the Warped DGP Brane}\\
\small \vspace{1cm}
{\bf Kourosh Nozari}\quad  and \quad{\bf Behnaz Fazlpour}\\
\vspace{0.5cm} {\it Department of Physics, Faculty of Basic
Sciences,\\
University of Mazandaran,\\
P. O. Box 47416-1467,
Babolsar, IRAN\\
e-mail: knozari@umz.ac.ir}

\end{center}
\vspace{1.5cm}
\begin{abstract}
We construct an inflation model with inflaton non-minimally coupled
to gravity on a warped DGP brane. Using an exponential potential, we
calculate scalar power spectrum, spectral index and the running of
the spectral index. We show that for a suitable range of non-minimal
coupling it is possible to exit the inflationary phase spontaneously
and without any mechanism, even in the case that minimal inflation
can not exit spontaneously. By a detailed analysis of parameter
space, we study the constraints imposed on the non-minimal coupling
from recent observational data.

\end{abstract}
\vspace{2cm}
\newpage
\section{Introduction}
Inflation is one of the most important achievements in modern
cosmology[1]. It has became the standard scenario for the early
universe since it has the potential to solve some outstanding
problems present in the standard Hot Big-Bang cosmology. More
importantly, it produces the cosmological fluctuations for the
formation of the structure that we observe today. Despite the great
successes of inflation paradigm, there are several serious problems
with no concrete solutions; natural realization of inflation in a
fundamental theory, cosmological constant and dark energy problem,
unexpected low power spectrum at large scales and egregious running
of spectral index are some of these problems( for a list of
theoretical problems in inflation paradigm, see [2]). Another
unsolved problem in the spirit of inflationary scenario is that we
do not know how to integrate it with ideas in particle physics. For
example, we would like to identify the inflaton, the scalar field
that drives inflation, with one of the known fields of particle
physics. Furthermore, it is important that the inflaton potential
emerges naturally from underling fundamental theory[3].

In the spirit of the braneworld scenarios, there are promising
evidences to overcome some of these difficulties. Brane models are
inspired from M/string theory and branes are topological solitons in
nonperturbative M/string theories in $10$ or $11$ dimensional
spacetime. As Horava and Witten have shown, gauge fields of the
standard model are confined on two $9$-branes located at the end
points of an $S_1/Z_2$ orbifold[4]. Based on the Horava-Witten
model, the idea that our universe is a 3-brane embedded in a higher
dimensional spacetime has received a great deal of attention in
recent years(see [5] and references therein). In the braneworld
scenario, the standard model particles are confined on the 3-brane,
while the gravitation can propagate in the whole space. Since string
theory claims to give us a fundamental description of the nature, it
is important to study what kind of cosmology it predicts.
Furthermore, despite the fact that inflationary models have been
analyzed in standard four-dimensional cosmology, it is challenging
to discuss them in alternative gravitational theories such as brane
gravity. Among various braneworld scenarios, the model proposed by
Dvali, Gabadadze and Porrati (DGP) [6] is different in this respect
that it predicts deviations from the standard $4$-dimensional
gravity even over large distances. In this scenario, the transition
between four and higher-dimensional gravitational potentials arises
due to the presence of both the brane and bulk Einstein terms in the
action. Existence of a higher dimensional embedding space allows for
the existence of bulk or brane matter which can certainly influence
the cosmological evolution on the brane. In the DGP model, the bulk
is a flat Minkowski spacetime, but a reduced gravity term appears on
the brane without tension. Maeda, Mizuno and Torii have constructed
a braneworld scenario which combines the Randall-Sundrum II ( RS II)
model and DGP model[7]. In this combination, an induced curvature
term appears on the brane in the RS II model. This model has been
called {\it warped DGP braneworld} in literature[8]. Braneworld
model with scalar field minimally or non-minimally coupled to
gravity have been studied extensively(see[9] and references
therein). The introduction of non-minimal coupling is not just a
matter of taste; it is forced upon us in many situations of physical
and cosmological interests such as quantum corrections to the scalar
field theory and its renormalizability in curved spacetime[10]. In
the spirit of braneworld inflation scenario, there are numerous
studies which we highlight some of these studies based on their
importance and relevance to present work. The chaotic inflation
model on the RS II brane has been studied by Maartens {\it et
al}[11]. The inflation model in the braneworld scenario with a
Gauss-Bonnet term in the bulk has been discussed by Lidsey and Nunes
[12]. Scalar perturbation from braneworld inflation has been studied
by Koyama {\it et al}[13]. Barnaby, Burgess and Cline have studied
warped reheating in brane-antibrane inflation[14]. Brane inflation
and the cosmic string tension in superstring theory has been studied
by Firouzjahi and Henry Tye[15]. Curvaton reheating mechanism in
inflation on the warped DGP brane has been studied by Zhang and
Zhu[16]. Huang, Li and She have used WMAP three years data to
constrain brane inflation models[17]. Very recently, Bean {\it et
al} have compared brane inflation with WMAP data[18]. Panotopoulos
has studied assisted chaotic inflation in braneworld cosmology[19].
In addition to these studies, there are several other extensive
studies in braneworld inflation which can be obtained in
literature(for a recent but uncomplete list see[19]). Here we are
going to investigate non-minimal inflation in warped DGP braneworld.
In this direction, inflation model with {\it minimally coupled}
scalar field on the warped DGP braneworld has been studied by Cai
and Zhang[8]. Recently, Non-minimal inflation and running of the
spectral index have been studied by Li in usual 4-dimensional
spacetime[20]. This study which is restricted to 4-dimensional
cosmology, has been extended to high derivative coupling by Chen
{\it et al}[21]. The issue of non-minimal inflation on a warped DGP
brane has not been studied yet and therefore our goal in this paper
is to do this end to fill the existing gap. We study an inflation
model on the warped DGP scenario where inflaton is non-minimally
coupled to induced Ricci scalar on the brane. We calculate
parameters of our inflation model to study implications and
predictions of the model. We assume inflation of the universe is
driven by a single scalar field non-minimally coupled to induced
gravity with exponential potential on the warped DGP brane. We show
that the inflationary phase can exit spontaneously by a suitable
choice of non-minimal coupling. As we will show, depending on the
values of non-minimal coupling, the running of the scalar spectral
index can take negative values, which is in agreement with high
precision observations of WMAP[22]. We study the constraints imposed
on non-minimal coupling from observational data. As Faraoni has
shown, with non-minimal coupling it is harder to achieve inflation
and accelerated expansion[10]. However, since inclusion of
non-minimal coupling is forced upon us from quantum field theory
considerations in curved space, it is interesting to study the
effect of non-minimally coupled inflaton on a warped DGP brane.

\section{Warped DGP Braneworld}
Consider a 5-dimensional bulk spacetime with a single 4-dimensional
brane, on which gravity is localized. We write the action of our
model as follows[7]
\begin{equation}
{\cal{S}}=\int_{bulk}d^{5}X\sqrt{-{}^{(5)}g}\bigg[\frac{1}{2\kappa_{5}^{2}}
{}^{(5)}R+{}^{(5)}{\cal{L}}_{m}\bigg]+\int_{brane}d^{4}x\sqrt{-g}\bigg[\frac{1}{\kappa_{5}^{2}}
K^{\pm}+{\cal{L}}_{brane}(g_{\alpha\beta},\psi)\bigg].
\end{equation}
In this action the quantities are defined as follows: $X^{A}$ with
$A=0,1,2,3,5$ are coordinates in bulk while $x^{\mu}$ with
$\mu=0,1,2,3$ are induced coordinates on the brane. $\kappa_{5}^{2}$
is 5-dimensional gravitational constant. ${}^{(5)}R$ and
${}^{(5)}{\cal{L}}_{m}$ are 5-dimensional Ricci scalar and matter
Lagrangian respectively. $K^{\pm}$ is trace of extrinsic curvature
on either side of the brane. This term is known as
York-Gibbons-Hawking term[23].
${\cal{L}}_{brane}(g_{\alpha\beta},\psi)$  is the effective
4-dimensional Lagrangian. This action is actually a combination of
Randall-Sundrum II model[24] and DGP model[25]. In other words, an
induced curvature term is appeared on the brane in Randall-Sundrum
II model. So, this model is called {\it Warped DGP Braneworld} [8].
Consider the brane Lagrangian as follows
\begin{equation}
{\cal{L}}_{brane}(g_{\alpha\beta},\psi)=\frac{\mu^2}{2}R-\lambda+L_{m}
\end{equation}
where $\mu$ is a mass parameter, $R$ is Ricci scalar of the brane,
$\lambda$ is tension of the brane and $L_{m}$ is Lagrangian of the
other matters localized on the brane. Assume that bulk contains only
a cosmological constant, $^{(5)}\Lambda$. With these choices, action
(1) gives either a generalized DGP or a generalized RS II model: it
gives DGP model if $\lambda=0$ and $^{(5)}\Lambda=0$ and gives RS II
model if $\mu=0$ [7].

Considering a flat FRW metric on the brane, the dynamical equation
of the brane is given by
\begin{equation}
H^{2}=\frac{1}{3\mu^2}\bigg[\rho+\rho_{0}\Big(1+\varepsilon
{\cal{A}}(\rho,a)\Big)\bigg],
\end{equation}
where $\varepsilon=\pm 1$ corresponding to two possible branches of
solutions in this warped DGP model and
${\cal{A}}=\bigg[{\cal{A}}_{0}^{2}+\frac{2\eta}{\rho_{0}}\Big(\rho-\mu^{2}\frac{{\cal{E}}_{0}}{a^{4}}\Big)\bigg]^{1/2}$
where \,\, ${\cal{A}}_{0}\equiv
\bigg[1-2\eta\frac{\mu^{2}\Lambda}{\rho_{0}}\bigg]^{1/2}$,\,\, $\eta
\equiv\frac{6m_{5}^{6}}{\rho_{0}\mu^{2}}$\,\, with $0<\eta\leq1$
\,\,and \,\,$\rho_{0}\equiv
m_{\lambda}^{4}+6\frac{m_{5}^{6}}{\mu^{2}}$. Note that
${\cal{E}}_{0}$ is an integration constant and corresponding term in
the generalized Friedmann equation is called dark radiation term.
Since we are interested in the inflation dynamics of our model, we
neglect dark radiation term in which follows. In this case,
generalized Friedmann equation (3) attains the following form
\begin{equation}
H^{2}=\frac{1}{3\mu^2}\bigg[\rho+\rho_{0}+\varepsilon\rho_{0}\Big({\cal{A}}_{0}^{2}+\frac{2\eta\rho}{\rho_{0}}\Big)^{1/2}\bigg].
\end{equation}
This equation is the basis of our forthcoming arguments.

\section{Non-Minimal Inflation}
Minimal inflation on a warped DGP brane has been studied by Cai and
Zhang [8]. Here we consider the case of non-minimal inflation in a
warped DGP braneworld scenario described in the previous section. We
assume that the inflation is driven by non-minimally coupled scalar
field, $\varphi$ with potential $V(\varphi)$ on the warped DGP
brane. The action of this non-minimally coupled scalar field is
given by the following relation
\begin{equation}
{\cal{S}}_{\varphi}=\int_{brane}d^{4}x\sqrt{-g}\Big[\frac{1}{2}\xi
R\varphi^{2}-\frac{1}{2}\partial_{\mu}\varphi\partial^{\mu}\varphi-V(\varphi)\Big]
\end{equation}
where $\xi $ is a non-minimal coupling and $R$ is Ricci scalar of
the brane. Variation of the action with respect to $\varphi$ gives
the equation of motion of the scalar field
\begin{equation}
\ddot{\varphi}+3H\dot{\varphi}-\xi R\varphi +\frac{dV}{d\varphi}=0.
\end{equation}
The energy density and pressure of the non-minimally coupled scalar
field are given by
\begin{equation}
\rho=\frac{1}{2}\dot{\varphi}^{2}+V(\varphi)-6\xi R \varphi H
\dot{\varphi},
\end{equation}
\begin{equation}
p=\frac{1}{2}\dot{\varphi}^{2}-V(\varphi)+2\big(\xi R
\varphi\ddot{\varphi}+2\xi R \varphi H \dot{\varphi}+\xi R
\dot{\varphi}^{2}\big).
\end{equation}
In the slow-roll approximation where $\dot{\varphi}^{2}\ll
V(\varphi)$ and $\ddot{\varphi}\ll|3H\dot{\varphi}|$, equation of
motion for scalar field and energy density take the following forms
respectively
\begin{equation}
3H\dot{\varphi}-\xi R\varphi +\frac{dV}{d\varphi}\approx 0,
\end{equation}
\begin{equation}
\rho\approx V(\varphi)-6\xi R \varphi H \dot{\varphi}.
\end{equation}
Now we define the following quantities
\begin{equation}
\epsilon=-\frac{\dot{H}}{H^{2}},
\end{equation}
\begin{equation}
\delta=\frac{1}{3H^{2}}\frac{d^{2}V}{d\varphi^{2}},
\end{equation}
\begin{equation}
\gamma^2=\frac{1}{(3H^{2})^{2}}\frac{d^{3}V}{d\varphi^{3}}\frac{dV}{d\varphi}.
\end{equation}
In the slow-roll approximation, using equations (4), (6) and (10) we
find
$$\epsilon=\frac{\mu^{2}}{2}\frac{1}{\rho^{2}}\Bigg[\Big(\frac{dV}{d\varphi}\Big)^{2}-\xi
R\frac{dV}{d\varphi}\Big(\varphi-\frac{dV}{d\varphi}-2\varphi\frac{d^{2}V}{d\varphi^2}\Big)-2\xi^2
R^2 \varphi\Big(3\frac{dV}{d\varphi}+\varphi\frac{d^{2}
V}{d\varphi^2}\Big)+4\xi^3 R^3 \varphi^2\Bigg]$$
\begin{equation}
\times\Bigg[\frac{1+\varepsilon
\eta\Big({\cal{A}}_{0}^{2}+\frac{2\eta\rho}{\rho_{0}}\Big)^{-1/2}}{\bigg(1+\frac{\rho_{0}}{\rho}
\Big[1+\varepsilon\Big({\cal{A}}_{0}^{2}+\frac{2\eta\rho}{\rho_{0}}\Big)^{1/2}\Big]\bigg)^{2}}\Bigg],
\end{equation}
where a part of the effects of non-minimal coupling is hidden in the
definition of energy density, $\rho$, which attains the following
form
\begin{equation}
\rho\approx V+2 \xi R \varphi \frac{dV}{d\varphi}-2\xi^2 R^2
\varphi^2.
\end{equation}
$\delta$ and $\gamma$ can be obtained similarly.

Now we consider an exponential potential defined as follows
\begin{equation}
V(\varphi)=V_{0}\exp{\Big(-\sqrt{\frac{2}{p}}\frac{\varphi}{\mu}\Big)},
\end{equation}
where $V_{0}$ and $p$ are constants. In the minimal case where
$\xi=0$, we have inflationary solution for $p>2$. However, inflation
does not take place for $p\leq 2$ in the case of single scalar
field. It has been shown that inflation can proceed for the case of
multi-scalar fields even if the individual scalar field has the
power less than unity(assisted inflation)[26].

In our non-minimal case and with this potential, the slow-roll
parameters become
$$\epsilon=\frac{1}{p}\frac{1}{(1-2\sqrt{\frac{2}{p}}\,\frac{\xi
R\varphi}{\mu})^2}\Bigg(1+\xi R\Big(\sqrt{\frac{p}{2}}\frac{\mu
\varphi}{V}+1-2\sqrt{\frac{2}{p}}\frac{\varphi}{\mu}\Big)\Bigg)$$
\begin{equation}
\times \Bigg[\frac{1+\varepsilon \eta\Big({\cal{A}}_{0}^{2}+2\eta
x\Big)^{-1/2}}{\bigg(1+\frac{1}{x}
\Big[1+\varepsilon\Big({\cal{A}}_{0}^{2}+2\eta
x\Big)^{1/2}\Big]\bigg)^{2}}\Bigg],
\end{equation}
\begin{equation}
\delta=\frac{2}{p}\,\,\frac{x}{\Big(1-2\sqrt{\frac{2}{p}}
\frac{\xi R
\varphi}{\mu}\Big)\Big[x+1+\varepsilon\Big({\cal{A}}_{0}^{2}+2\eta
x\Big)^{1/2}\Big]}
\end{equation}
where $x=\frac{\rho}{\rho_{0}}$ and
\begin{equation}
\gamma^{2}=\delta^{2}.
\end{equation}
Only terms of $\cal{O}(\xi)$ have been considered in these and
forthcoming calculations. Substituting (16) into (15), $\rho$ takes
the following form
\begin{equation}
\rho \approx V\Big(1-2\sqrt{\frac{2}{p}} \frac{\xi R
\varphi}{\mu}\Big)
\end{equation}
For comparison, we briefly discuss the minimal case where $\xi=0$.
In this case the inflation will ends when $\epsilon=1$. The Universe
enters inflationary phase where $p>2$. In this minimal case Universe
can not exit inflationary phase in the branch of $\varepsilon=1$
while in the branch $\varepsilon=-1$ it can exit inflationary phase
naturally without any other mechanism[8]. However, in our
non-minimal case the situation is very different. Existence of
non-minimal coupling provides the situation that Universe can exit
inflationary phase even in branch $\varepsilon=1$. To show this end,
we should solve the following complicated equations
$$\frac{1}{p}\frac{1}{(1-2\sqrt{\frac{2}{p}}\,\frac{\xi
R\varphi}{\mu})^2}\Bigg(1+\xi R\Big(\sqrt{\frac{p}{2}}\frac{\mu
\varphi}{V}+1-2\sqrt{\frac{2}{p}}\frac{\varphi}{\mu}\Big)\Bigg)$$
\begin{equation}
\times \Bigg[\frac{1+\eta\Big({\cal{A}}_{0}^{2}+2\eta
x\Big)^{-1/2}}{\bigg(1+\frac{1}{x}
\Big[1+\Big({\cal{A}}_{0}^{2}+2\eta
x\Big)^{1/2}\Big]\bigg)^{2}}\Bigg]=1,
\end{equation}
for $\varepsilon=1$, and
$$\frac{1}{p}\frac{1}{(1-2\sqrt{\frac{2}{p}}\,\frac{\xi
R\varphi}{\mu})^2}\Bigg(1+\xi R\Big(\sqrt{\frac{p}{2}}\frac{\mu
\varphi}{V}+1-2\sqrt{\frac{2}{p}}\frac{\varphi}{\mu}\Big)\Bigg)$$
\begin{equation}
\times \Bigg[\frac{1-\eta\Big({\cal{A}}_{0}^{2}+2\eta
x\Big)^{-1/2}}{\bigg(1+\frac{1}{x}
\Big[1-\Big({\cal{A}}_{0}^{2}+2\eta
x\Big)^{1/2}\Big]\bigg)^{2}}\Bigg]=1,
\end{equation}
for $\varepsilon=-1$. These equations can not be solved in algebraic
way, so we try to solve them numerically. As figure $1$ shows, for
$\varepsilon=-1$ inflationary phase can exit without any mechanism
for positive values of non-minimal coupling. Also for some small
negative values of non-minimal coupling such as $\xi=-\frac{1}{12}$
this spontaneous exit is possible. However, some values of
non-minimal coupling such as $\xi=-\frac{1}{6}$ prevents spontaneous
exit of inflation. In fact, for all $\xi <-\frac{1}{6}$ this
spontaneous exit fails in branch $\varepsilon=-1$. In the case of
$\varepsilon=+1$ inflationary phase can not exit spontaneously with
small (positive and negative) values of non-minimal coupling(figure
$2$). However, for relatively large values of non-minimal coupling,
(for instance $\xi\geq1.55$) it is possible to exit inflationary
phase spontaneously even in $\varepsilon=+1$ branch. This is the
main deference of our non-minimal framework with minimal case [8].
Note that in the original DGP model, spontaneous exit of inflation
is possible but inflationary phase lasts to extremely low energy
scale ($\rho_{0}\sim (10^{-3}eV)^{4}$) which is evidently
impossible. In our non-minimal framework this difficulty can be
avoided by choosing a suitable range of non-minimal coupling.

The number of e-folds, $N\equiv\ln\frac{a_{e}}{a_{i}}$ can be
written as
\begin{equation}
N=-\int_{\varphi_{i}}^{\varphi_{e}}3H^{2}\frac{d\varphi}{dV}d\varphi,
\end{equation}
where $\varphi_{i}$ denotes the value of scalar field $\varphi$
when Universe scale observed today crosses the Hubble horizon
during inflation, while $\varphi_{e}$ is the value of scalar field
when the Universe exits the inflationary phase. In the presence of
non-minimal coupling and with exponential potential (16), the
number of e-folds is given by
\begin{equation}
N=\frac{1}{\mu}\,\sqrt{\frac{p}{2}}\int_{\varphi_{i}}^{\varphi_{e}}\frac{\Big(1-2\sqrt{\frac{2}{p}}\frac{\xi
R \varphi}{\mu}\Big)}{x}\,\Big[x+1+\varepsilon \omega\Big]d\varphi,
\end{equation}
where $\omega=\Big({\cal{A}}_{0}^{2}+2\eta x\Big)^{1/2}$ \, and \,
$x=\frac{V\Big(1-2\sqrt{\frac{2}{p}}\frac{\xi R
\varphi}{\mu}\Big)}{\rho_{0}}$ \, so that \, $d\varphi=
-\frac{dx}{\sqrt{\frac{2}{p}}\frac{1}{\mu}(x+2\frac{\xi R
V}{\rho_{0}})}$\,.\, $x_{i}$ is the value of $x$ when the cosmic
scale observed today crosses the Hubble horizon during inflation.

In the next step we consider the scalar perturbation of the metric.
These perturbations are supposed to be adiabatic. We define the
scalar spectrum index as follows
\begin{equation}
n_{s}=1+\frac{d\ln A_{s}^{2}}{d\ln k},
\end{equation}
where $A_{s}$ is the scalar curvature perturbation amplitude of a
given mode when re-enters the Hubble radius, defined as
$A_{s}^{2}=\frac{H^{4}}{(25\pi^{2}\dot{\varphi}^2)}$. Substituting
(9) in this relation, we obtain
\begin{equation}
A_{s}^{2}=\Bigg[\frac{9H^{6}}{25\pi^{2}\Big(\xi R
\varphi-\frac{dV}{d\varphi}\Big)^{2}}\Bigg]_{k=aH},
\end{equation}
Note that although the non-minimal coupling of the scalar field to
the Ricci curvature on the brane leads to the non-conservation of
the scalar field effective energy density[27,28]
\begin{equation}
\dot{\rho}+3H(\rho+p)=-6\xi R\varphi\dot{\varphi}(H^{2}+\dot{H}),
\end{equation}
since the total energy-momentum on the brane is conserved, the
curvature perturbation on a uniform density hypersurface is still
conserved on large scale. Using parameters of slow-rolling, we have
\begin{equation}
n_{s}=1-6\,\epsilon+2\delta-\frac{2}{3}\frac{\xi R}{H^2}.
\end{equation}
The running of the spectral index is defined as follows
\begin{equation}
\frac{dn_{s}}{d\ln
k}=16\epsilon\delta-24\epsilon^{2}+4\gamma^{2}-\frac{16}{3}\frac{\xi
R}{H^2}\,\epsilon-\frac{4\xi R
\varphi}{\frac{dV}{d\varphi}}\,\gamma^{2}.
\end{equation}
For comparison purposes, we first write the explicit form of these
quantities in the minimal framework where $\xi =0$\,\,[8]
\begin{equation}
n_{s}=1-\frac{2}{p}\,\,\frac{x' \Big(2{\cal{A}}_{0}^{2}+\eta
x'+(-2+x')\omega \Big)}{\omega(1+x'-\omega)^2},\hspace{10cm}
\end{equation}
and
$$\frac{dn_{s}}{d\ln
k}=\frac{8}{p^{2}}\,\,\frac{x'\,^{2}}{\omega^{2}(1+x'-\omega)^{4}}\times\hspace{11cm}$$
\begin{equation}
\Big[2{\cal{A}}_{0}^{4}+{\cal{A}}_{0}^{2}\Big(2+8x'-4(1+2x')\omega+3x'\,^2+12
\eta x'\Big)+\eta x'\Big(4+x'(16+13\eta-14\omega)+6x'\,^2-12
\omega\Big)\Big],\hspace{2cm}
\end{equation}
where $x'=\frac{V}{\rho_{0}}$. However, in the presence of
non-minimal coupling these quantities attain the following more
complicated forms
$$n_{s}=1-\frac{2}{p}\,\,\frac{x}{(1-2\sqrt{\frac{2}{p}}\,\frac{\xi
R\varphi}{\mu})(1+x+\varepsilon \omega)}\times\hspace{10cm}$$
\begin{equation}
\Bigg[-2+\frac{\mu^{2}\xi R p}{V}+\frac{3 x\,(1+\varepsilon \eta
\omega^{-1})}{(1-2\sqrt{\frac{2}{p}}\,\frac{\xi
R\varphi}{\mu})(1+x+\varepsilon \omega)}\Bigg(1+\xi
R\Big(\sqrt{\frac{p}{2}}\frac{\mu
\varphi}{V}+1-2\sqrt{\frac{2}{p}}\frac{\varphi}{\mu}\Big)\Bigg)\Bigg],\hspace{2cm}
\end{equation}
and
$$\frac{dn_{s}}{d\ln
k}=\frac{8}{p^{2}(1-2\sqrt{\frac{2}{p}}\,\frac{\xi
R\varphi}{\mu})^2}\,\,\Big(\frac{x}{x+1+\varepsilon\omega}\Big)^2\times\hspace{11cm}$$

$$\Bigg[2(1+\sqrt{\frac{p}{2}}\frac{\xi R\varphi\mu}{V})+\Big(4-\frac{2\xi R
\mu^{2}p}{V}\Big)\Big(\frac{x}{x+1+\varepsilon\omega}\Big)(1+\varepsilon\eta\omega^{-1})\Bigg(\frac{1+\xi
R\Big(\sqrt{\frac{p}{2}}\,\frac{\mu\varphi}{V}+1-2\sqrt{\frac{2}{p}}\,\frac{\varphi}{\mu}\Big)}{1-2\sqrt{\frac{2}
{p}}\,\frac{\xi R \varphi}{\mu}}\Bigg)\hspace{3cm}$$
\begin{equation}
-3\Big(\frac{x}{x+1+\varepsilon\omega}\Big)^{2}(1+\varepsilon\eta\omega^{-1})^{2}\Bigg(\frac{1+\xi
R\Big(\sqrt{\frac{p}{2}}\,\frac{\mu\varphi}{V}+1-2\sqrt{\frac{2}{p}}\,\frac{\varphi}{\mu}\Big)}{1-2\sqrt{\frac{2}
{p}}\,\frac{\xi R \varphi}{\mu}}\Bigg)^{2}\Bigg],\hspace{5cm}
\end{equation}
where as usual \, $x=\frac{V \Big(1-2\sqrt{\frac{2}{p}}\,\frac{\xi R
\varphi}{\mu}\Big)}{\rho_{0}}$. The relation between $x$ and $x'$ is
\begin{equation}
x=x'\Big(1-2\sqrt{\frac{2}{p}}\,\frac{\xi R \varphi}{\mu}\Big),
\end{equation}
The COBE normalization gives us $A_{s}^2=4\times10^{-10}$, therefore
equation (26) leads to
\begin{equation}
\Bigg[\frac{\Big[V\Big(1-2\sqrt{\frac{2}{p}} \frac{\xi R
\varphi}{\mu}\Big)+\rho_{0}+\varepsilon
\rho_{0}\Big({\cal{A}}_{0}^{2}+\frac{2\eta\rho}{\rho_{0}}\Big)^{1/2}\Big]^3}{75
\pi^2 \mu^6 \Big(\xi R \varphi+\sqrt{\frac{2}{p}}\,\frac{1}{\mu}
V\Big)^{2}} \Bigg]_{k=aH}=4 \times 10^{-10}.
\end{equation}
This relation can be rewritten as follows
\begin{equation}
\frac{\Big(1-2\sqrt{\frac{2}{p}} \frac{\xi R
\varphi}{\mu}\Big)}{\Big(1+\sqrt{\frac{p}{2}}\,\frac{\xi
R\varphi\mu }{V}\Big)^{\frac{2}{3}}}\times
\Big[1+\frac{1}{x}+\frac{\varepsilon}{x}({\cal{A}}_{0}^{2}+2\eta
x)^{\frac{1}{2}}\Big]=B
\end{equation}
where \, $B=(\frac{6\times 10^{-7}}{y p})^{\frac{1}{3}}$\,\,  and
\, \,$y=\frac{V_{i}}{\mu^4}$\,. Here the subscript $i$ means that
the corresponding quantity should be calculated at $k=a H$. From
equation (36) we get
\begin{equation}
x_{i}=\frac{2(-1+D+\eta)}{(-1+D)^2},
\end{equation}
where
\begin{equation}
D=B\times\frac{\Big(1+\sqrt{\frac{p}{2}}\,\frac{\xi R\varphi\mu
}{V}\Big)^{\frac{2}{3}}}{\Big(1-2\sqrt{\frac{2}{p}} \frac{\xi R
\varphi}{\mu}\Big)}.
\end{equation}
In the minimal limit \,$\xi=0$, when $y=y'_{s}=\frac{6\times
10^{-7}}{p}$, there is a singularity that appears when
$x_{i}\rightarrow\infty$ in (37) since in this case \, $D=1$. This
singularity restricts the energy scale of the inflation in such a
way that $y<y'_{s}$ should be satisfied. We can see that with a
non-minimally coupled inflaton, the restriction on the energy scale
of the inflation is a function of non-minimal coupling $\xi$ as
follows
\begin{equation}
y<y_{s},
\end{equation}
where energy scale $y_{s}$ has the following complicated non-minimal
coupling dependence
\begin{equation}
y_{s}=(\frac{6\times 10^{-7}}{p})\times
\frac{\Big(1+\sqrt{\frac{p}{2}}\,\frac{\xi R\varphi\mu
}{V}\Big)^{2}}{\Big(1-2\sqrt{\frac{2}{p}} \frac{\xi R
\varphi}{\mu}\Big)^3}.
\end{equation}
If $y>y_{s}$, we have $x_{i}=\infty$ or $\varphi_{i}=\infty$. This
leads to a negative number of e-folds and therefore an unphysical
result. The energy scale of inflation for different values of
non-minimal coupling will be discussed in the next section.\\
Before a detailed analysis of parameter space, we should stress that
with non-minimal coupling of inflaton and curvature, one can perform
a conformal transformation to a new metric for which the coupling of
the inflaton to the metric becomes minimal(see for example [10,
30-33]). In the new frame(Einstein frame), the inflaton potential
gets an additional term. As Komatsu and Futamase have shown[30],
generally slow-roll parameters are different in these two frames
(Jordan and Einstein Frame). Spectral index and other inflationary
parameters are different in this frames also and only in a first
order theory these quantities are the same in two frames. In our
case, with non-minimally coupled inflaton on the warped DGP brane,
slow-roll parameters are shown to be very different with respect to
minimal case. In this case warp factor itself obtains a complicated
dependence on the non-minimal coupling. To see this end, remember
that ${\cal{A}}_{0}\equiv
\bigg[1-2\eta\frac{\mu^{2}\Lambda}{\rho_{0}}\bigg]^{1/2}$. With
non-minimally coupled inflaton, we should redefine $\eta$ as follows
\begin{equation}
\eta_{new}=\eta \Big(1-2\sqrt{\frac{2}{p}}\frac{\xi R
\varphi}{\mu}\Big).
\end{equation}
Now we may define a new ${\cal{A}}_{0}$ as follows
\begin{equation}
\Big({\cal{A}}_{0}\Big)_{new}\equiv \bigg[1-2\eta
\Big(1-2\sqrt{\frac{2}{p}}\frac{\xi R
\varphi}{\mu}\Big)\frac{\mu^{2}\Lambda}{\rho_{0}}\bigg]^{1/2}.
\end{equation}
This equation shows an explicit dependence of warp factor to the
non-minimal coupling of inflaton and Ricci scalar. As a result, with
non-minimally coupled inflaton, we are not faced only with a
redefinition of potential; we have a redefinition of slow-rolling
parameters also. In this case physical observable such as spectral
index attain a new and complicated form depending on both warp
factor and non-minimal coupling as follows
$$n_{s}=1-\frac{2}{p}\,\,\frac{x}{\omega(1-2\sqrt{\frac{2}{p}}\,\frac{\xi
R\varphi}{\mu})(1+x+\varepsilon \omega)^{2}}\times\hspace{10cm}$$
$$\Bigg[\omega \Bigg(-2-2x+3x\Bigg(1+\xi
R\Big(\sqrt{\frac{p}{2}}\frac{\mu
\varphi}{V}+1-2\sqrt{\frac{2}{p}}\frac{\varphi}{\mu}\Big)\Bigg)\Bigg)+\frac{\mu^{2}\xi
R p}{V}\Big(\omega(1+x)+\varepsilon({\cal{A}}_{0}^{2}+2\eta
x)\Big)-$$
\begin{equation}
\varepsilon \Bigg(2{\cal{A}}_{0}^{2}+4\eta x-3\eta x\Bigg(1+\xi
R\Big(\sqrt{\frac{p}{2}}\frac{\mu
\varphi}{V}+1-2\sqrt{\frac{2}{p}}\frac{\varphi}{\mu}\Big)\Bigg)\Bigg)\Bigg]\hspace{5cm}
\end{equation}
In fact as has been shown in [30], only for $\xi\gg 1$ physical
observables of inflation do not depend on $\xi$. Therefore, in our
case both warp factor and observable quantities are different
relative to minimal case and a frame (conformal) transformation
alone cannot relate our results to the results of minimal case
studied in [8].\\
With this point in mind, in the next section we present a detailed
analysis of parameter space. In which follows, in all numerical
calculations and resulting figures we have set
$\mu=\varphi=R={\cal{A}}_{0}=V_{0}=1$,\, $p=50$ and $\eta=0.99$.

\section{Numerical Analysis of the Parameter Space}
To explore cosmological implications of this non-minimal inflation
model, we perform some numerical analysis of parameter space. The
results of this analysis are shown in figures. Figure $1$ shows the
variation of the slow roll parameter $\epsilon$ with respect to $x$
for different values of non-minimal coupling on the branch
$\varepsilon=-1$ of the warped DGP braneworld. In minimal case,
inflation on this branch can exit spontaneously without any
mechanism since the equation $\epsilon=1$ has solution always. As
this figure shows, the situation for non-minimal case is different
in some respects. For example, if $\xi=-\frac{1}{6}$, it is
impossible to reach $\epsilon=1$ and therefore, there is no
spontaneous exit from inflation with this non-minimal coupling
within our framework. For negative larger values of non-minimal
coupling the situation is similar. Therefore, in contrast to minimal
inflation on the warped DGP brane, non-minimal inflation prevents
spontaneous exit from inflation on the $\varepsilon=-1$ branch for
some values of non-minimal coupling. Note that in this figure all
upper curves intersect $\epsilon=1$ line, but the lower curve
corresponding to $\xi=-\frac{1}{6}$ will not reach $\epsilon=1$ at
all. Figure $2$ shows the slow roll parameter $\epsilon$ versus
$x=\frac{V \Big(1-2\sqrt{\frac{2}{p}}\,\frac{\xi R
\varphi}{\mu}\Big)}{\rho_{0}}$ for different values of non-minimal
coupling on the branch $\varepsilon=+1$ of the warped DGP
braneworld. As this figure shows, for small values of non-minimal
coupling it is impossible to exit inflationary phase spontaneously
and within reasonable time scale. However, for larger values of
non-minimal coupling, in contrast to minimal case[8], it is possible
to exit from inflation spontaneously in branch $\varepsilon=+1$ with
some values of the non-minimal coupling in a reasonable time scale.
For instance, as figure $3$ shows, if we set $\xi=1.55$ or larger,
it is possible to reach $\epsilon=1$ in this branch. This is one of
the main differences between our model and the minimal model studied
in [8]. Figure $4$ shows the scalar spectrum index with
$\varepsilon=-1$. In minimal case, this index is always less than
unity(red spectrum)[8]. We see that for negative non-minimal
coupling it is possible to find scalar spectrum index larger than
unity(a blue spectrum). The result of WMAP3 for $\Lambda$CDM gives
$n_{s}=0.951^{+0.015}_{-0.019}$ for index of the power spectrum[29].
Combining WMAP3 with SDSS (Sloan Digital Sky Survey), gives
$n_{s}=0.948^{+0.015}_{-0.018}$ at the level of one standard
deviation[17]. These results show that a red power spectrum is
favored at least at the level of two standard deviations. If there
is running of the spectral index, the constraints on the spectral
index and its running are given by
\begin{equation}
n_{s}=1.21^{+0.13}_{-0.16}
\end{equation}
and
\begin{equation}
\frac{dn_{s}}{d\ln k}=-0.102^{+0.050}_{-0.043}
\end{equation}

We see from figure $4$ that in our non-minimal case, for some
negative values of non-minimal coupling it is possible to have
spectral index such that $n_{s}>1$. Therefore our non-minimal
inflation model is consistent with WMAP3 data with running of
spectral index. In other words, non-minimal inflation can be
supported by WMAP3 data with running of spectral index. The
situation for $\varepsilon=+1$ is similar. The main characteristic
of our setting in this regard is the fact that now $n_{s}>1$ is
possible for some values of non-minimal coupling. This situation is
impossible in minimal inflation on the warp DGP brane. On the other
hand, the running of the spectral index in minimal case is always
negative. In our non-minimal case, as figure $6$ with
$\varepsilon=-1$ shows, for small negative values of non-minimal
coupling (for instance, $\xi=-\frac{1}{12}$), it is possible to have
positive running of the spectral index. Therefore, in contrast to
minimal case, for a suitable range of non-minimal coupling running
of spectral index can attain large positive values. The situation
for branch $\varepsilon=+1$ is more or less similar (figure $7$),
however in this branch the running of the scalar spectrum index for
conformal coupling is approximately zero.

Now we investigate the energy scale of this non-minimal inflation.
We set $\eta=0.99$ and calculate the energy scales for different
values of non-minimal coupling. Firstly we should obtain $V_{i}$'s
using $y=\frac{V_{i}}{\mu^{4}}$ for different values of non-minimal
coupling. Considering $\rho \approx V\Big(1-2\sqrt{\frac{2}{p}}
\frac{\xi R \varphi}{\mu}\Big)$, we obtain the energy scales of this
non-minimal inflation as presented in table 1.\\
\begin{table}
\begin{center}
\caption{Energy Scales of Non-minimal Inflation} \vspace{0.5 cm}
\begin{tabular}{|c|c|c|c|c|c|c|c|}
  \hline
  \hline& & $V_{i}$ & $\rho_{i}$  \\
  \hline $\xi=\frac{1}{6}$ &$y\sim10^{-8}$ &$(10^{15}GeV)^4$ &$(0.98\times10^{15}GeV)^4$   \\
 \hline $\xi=\frac{1}{12}$& $y\sim10^{-8}$ & $(10^{15}GeV)^4$  & $(0.99\times10^{15}GeV)^4$ \\
  \hline$\xi=-\frac{1}{6}$& $y\sim10^{-12}$  & $(10^{14}GeV)^4$  & $(1.01\times10^{14}GeV)^4$ \\
  \hline$\xi=-\frac{1}{12}$&$y\sim10^{-9}$&$(5.62\times10^{14}GeV)^4$&$(5.66\times10^{14}GeV)^4$\\
\hline
\end{tabular}
\end{center}
\end{table}
For comparison, note that in the minimal case we have
$\rho_{i}\sim(5.62\times10^{14}GeV)^4$, therefore except for
$\xi=-\frac{1}{6}$, the non-minimal coupling leads to larger values
of energy scales of inflation. Figures $8$, $9$, $10$ and $11$ show
the non-minimal inflation energy scale versus the parameters $x$ and
$p$ for different values of non-minimal coupling. Increasing the
values that a positive non-minimal coupling can attains leads to the
larger inflation energy scale. The situation for negative
non-minimal coupling is similar. Since based on WMAP3 data, $
n_{s}=1.21^{+0.13}_{-0.16}$, or $1.05\leq n_{s}\leq1.34$, using
equation (32), (34)  and with $\rho_{\circ}\sim(10^{-3}eV)^{4}$, the
constraint on the non-minimal coupling from WMAP3 data are\,
$\xi\leq-1.05\times10^{-1}$\, and\,
$\xi\geq 1.2\times10^{-2}$.\\
In summary, inclusion of non-minimal coupling in inflation paradigm
is forced upon us from quantum field theory considerations in curved
space. In this paper we have studied non-minimal inflation on a
warped DGP braneworld. We have presented full dynamics of slow-roll
equations with non-minimally coupled scalar field and consistency of
the numerical results with observational data from WMAP3. Our model
allows for spontaneous exit from inflationary phase with some values
of non-minimal coupling and without any mechanism. Even for
situations that minimal case can not provide spontaneous exit,
non-minimal coupling can be chosen suitably to exit inflationary
phase spontaneously. This non-minimal model provides scalar spectrum
index larger than unity and also large positive values of running of
the spectral index. Finally, energy scale of inflation with
non-minimally coupled inflaton is larger than minimal case(except
for $\xi=-\frac{1}{6}$).\\

{\bf Acknowledgement}\\
A part of this work has been done during KN sabbatical leave at
Durham University, UK. He would like to appreciate members of the
Centre for Particle Theory of Durham University, specially Professor
Ruth Gregory, for kind hospitality. We would also to appreciate a
referee for His/Her valuable comments. \\

\begin{figure}[htp]
\begin{center}
\includegraphics{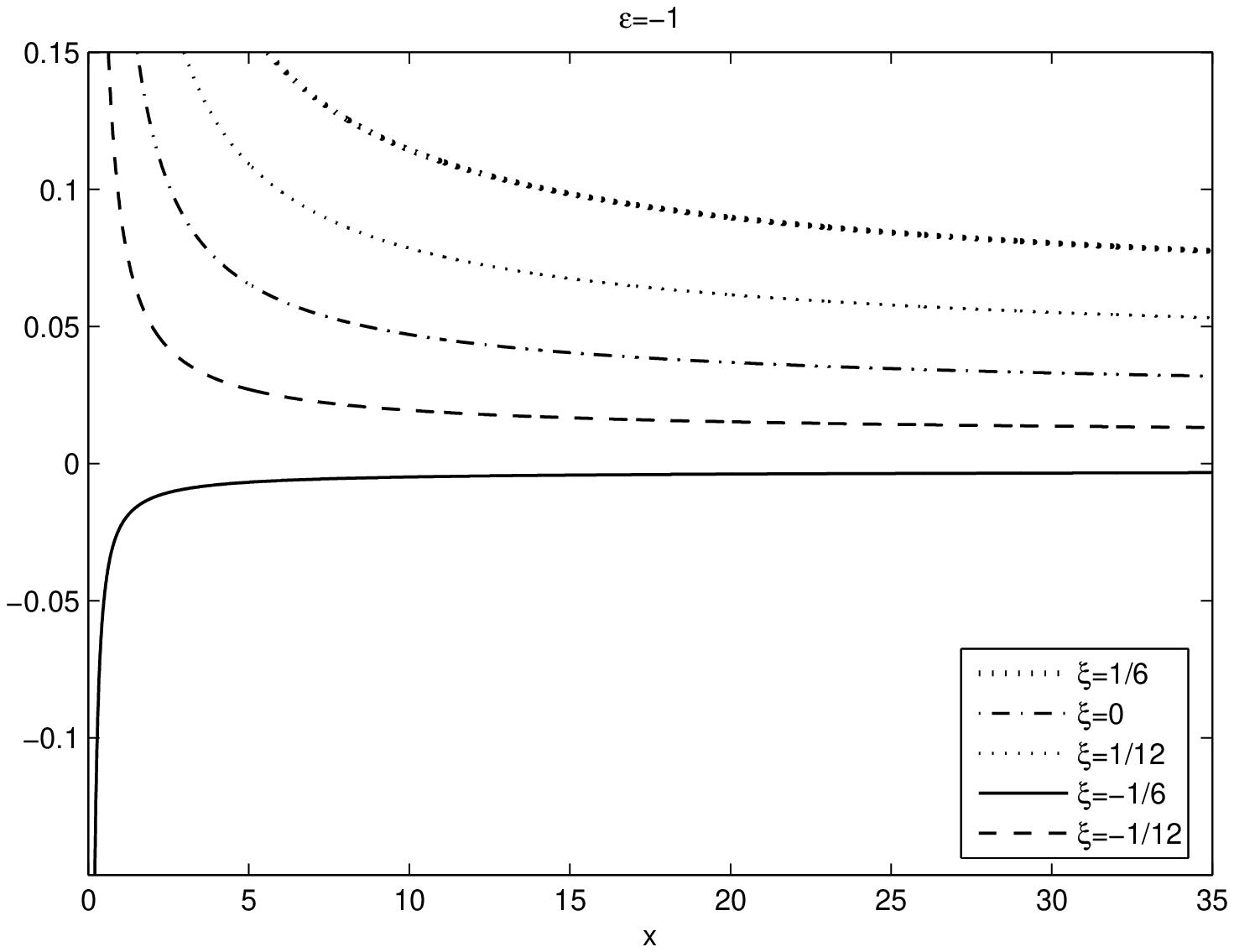}
\end{center}
\vspace{16 cm}
 \caption{\small { Variation of $\epsilon$ for different values
 of non-minimal coupling for branch $\varepsilon=-1$.}}
 \label{fig:1}
\end{figure}

\begin{figure}[htp]
\begin{center}
\includegraphics{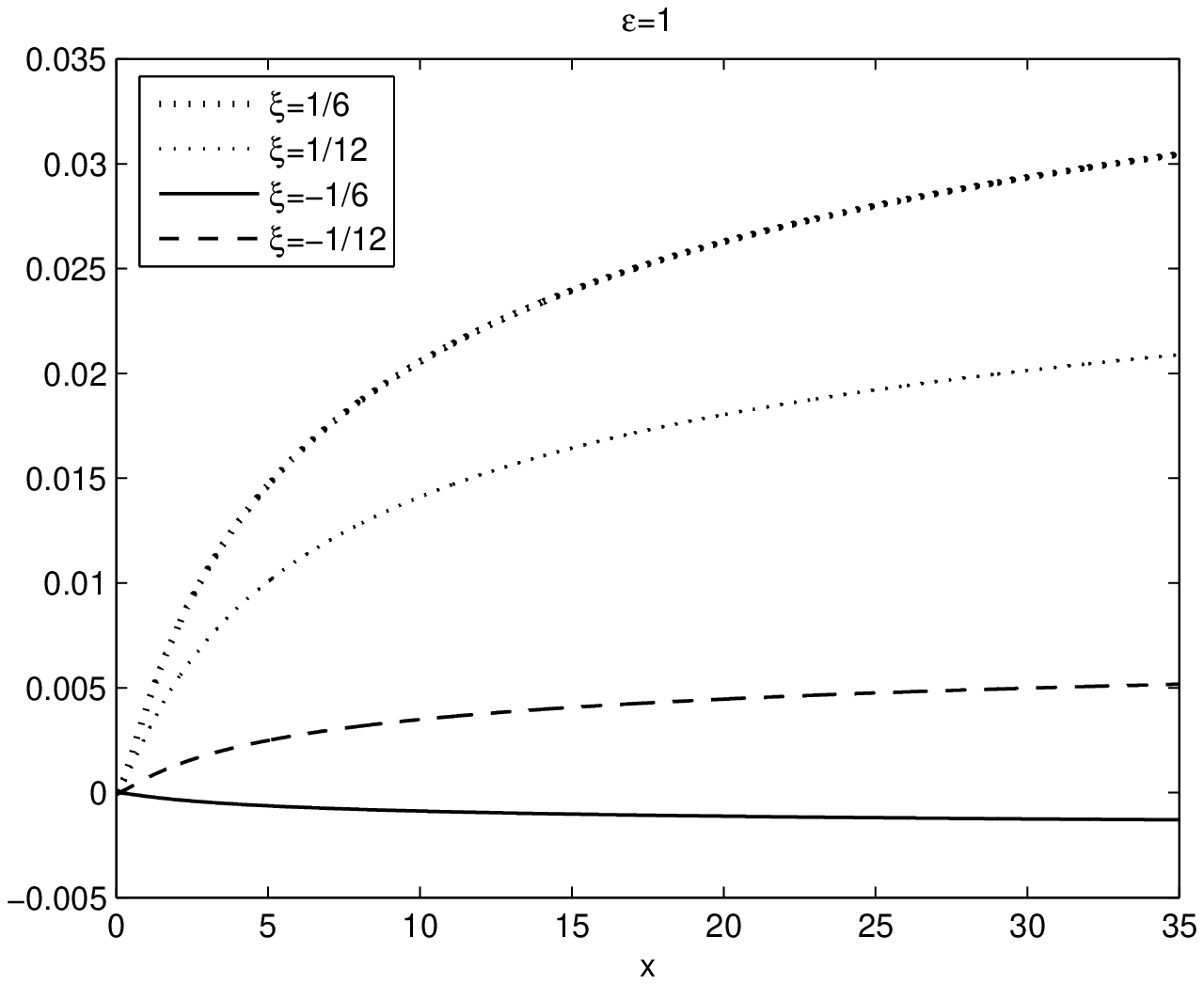}
\end{center}
\vspace{5 cm}
 \caption{\small { Variation of $\epsilon$ for different values
 of non-minimal coupling for branch $\varepsilon=+1$.}}
 \label{fig:2}
\end{figure}

\begin{figure}[htp]
\begin{center}
\includegraphics{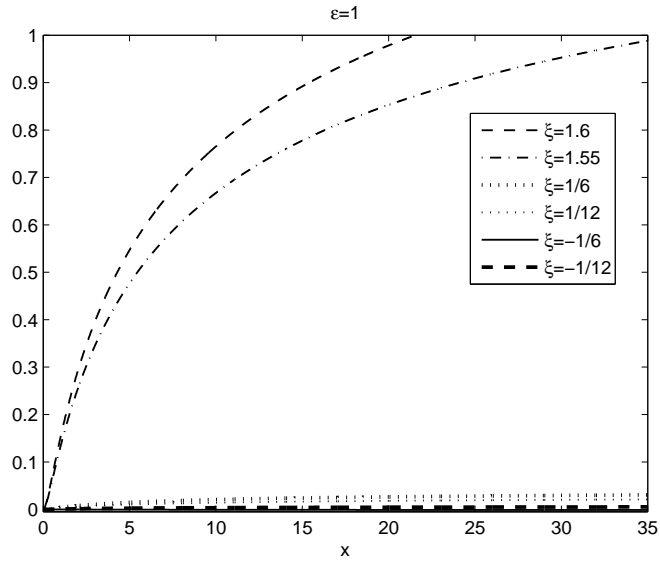}
\end{center}
\vspace{5 cm}
 \caption{\small { Variation of $\epsilon$ for different values
 of non-minimal coupling for branch $\varepsilon=+1$.
 For $\xi > 1.55$ it is possible to exit inflationary phase
 spontaneously without any mechanism.}}
 \label{fig:3}
\end{figure}

\begin{figure}[htp]
\begin{center}
\includegraphics{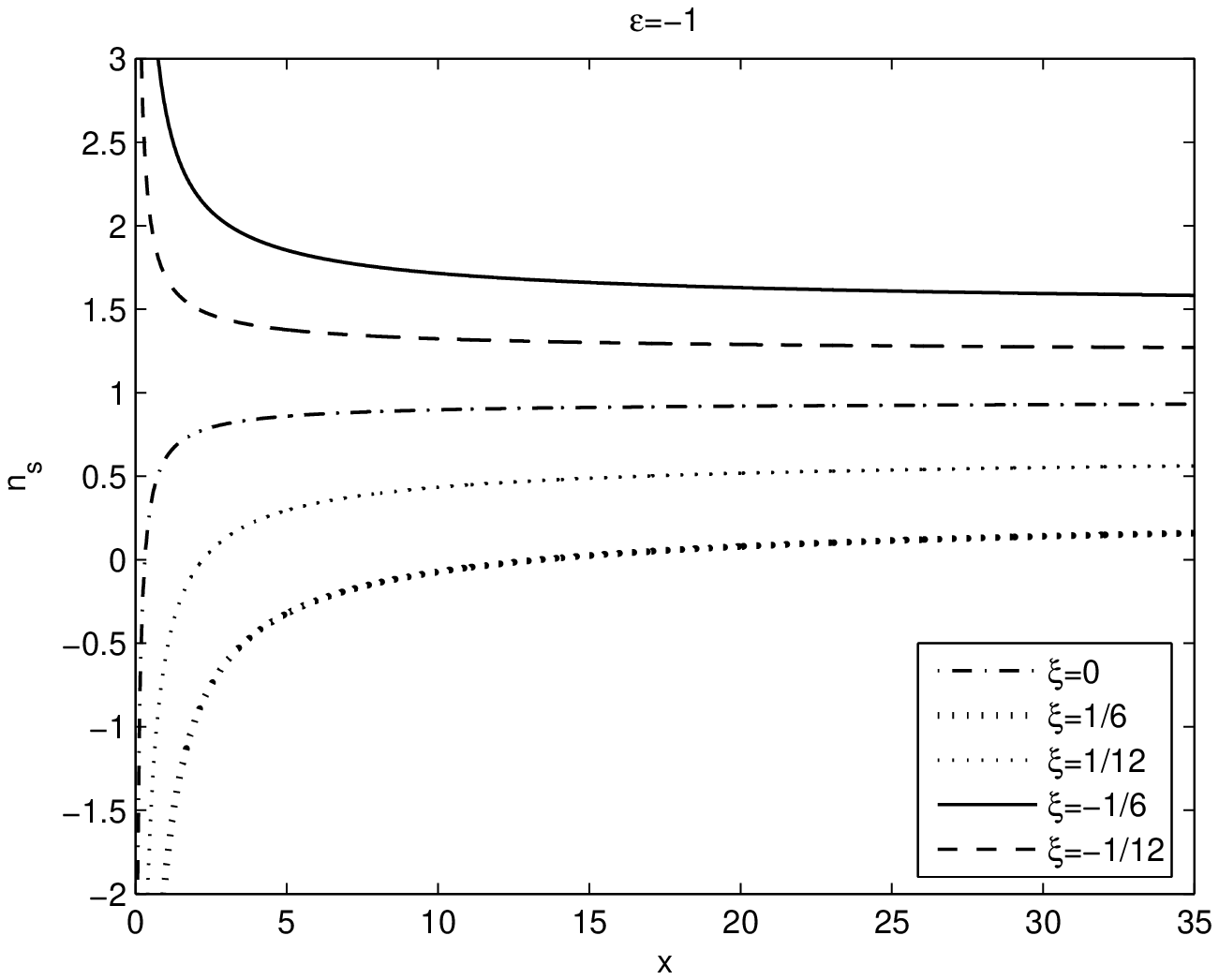}
\end{center}
\vspace{5 cm}
 \caption{\small { Spectral index $n_s$ for branch $\varepsilon=-1$.}}
 \label{fig:4}
\end{figure}

\begin{figure}[htp]
\begin{center}
\includegraphics{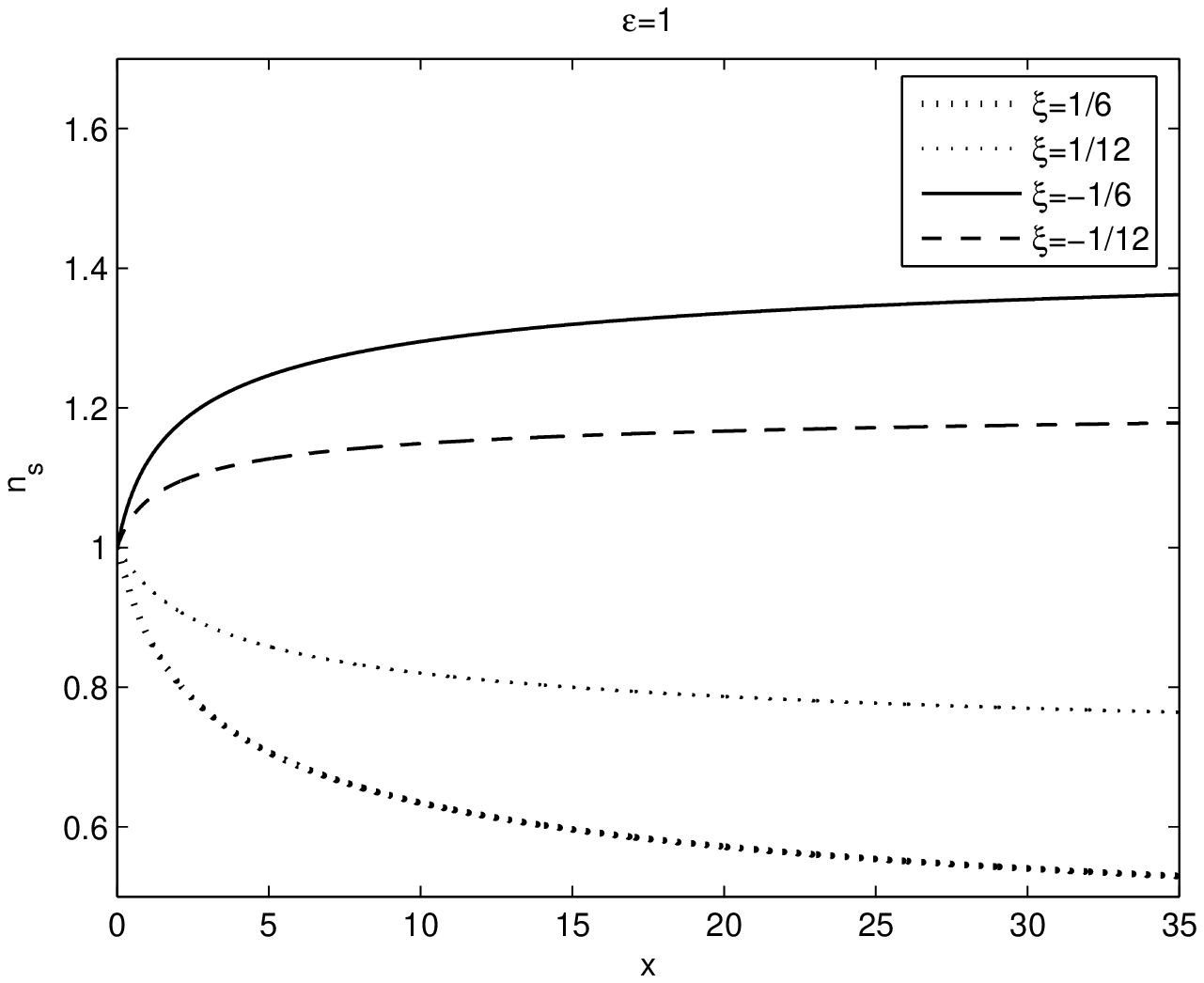}
\end{center}
\vspace{5 cm}
 \caption{\small { Spectral index $n_s$ for branch $\varepsilon=+1$.}}
 \label{fig:5}
\end{figure}

\begin{figure}[htp]
\begin{center}
\includegraphics{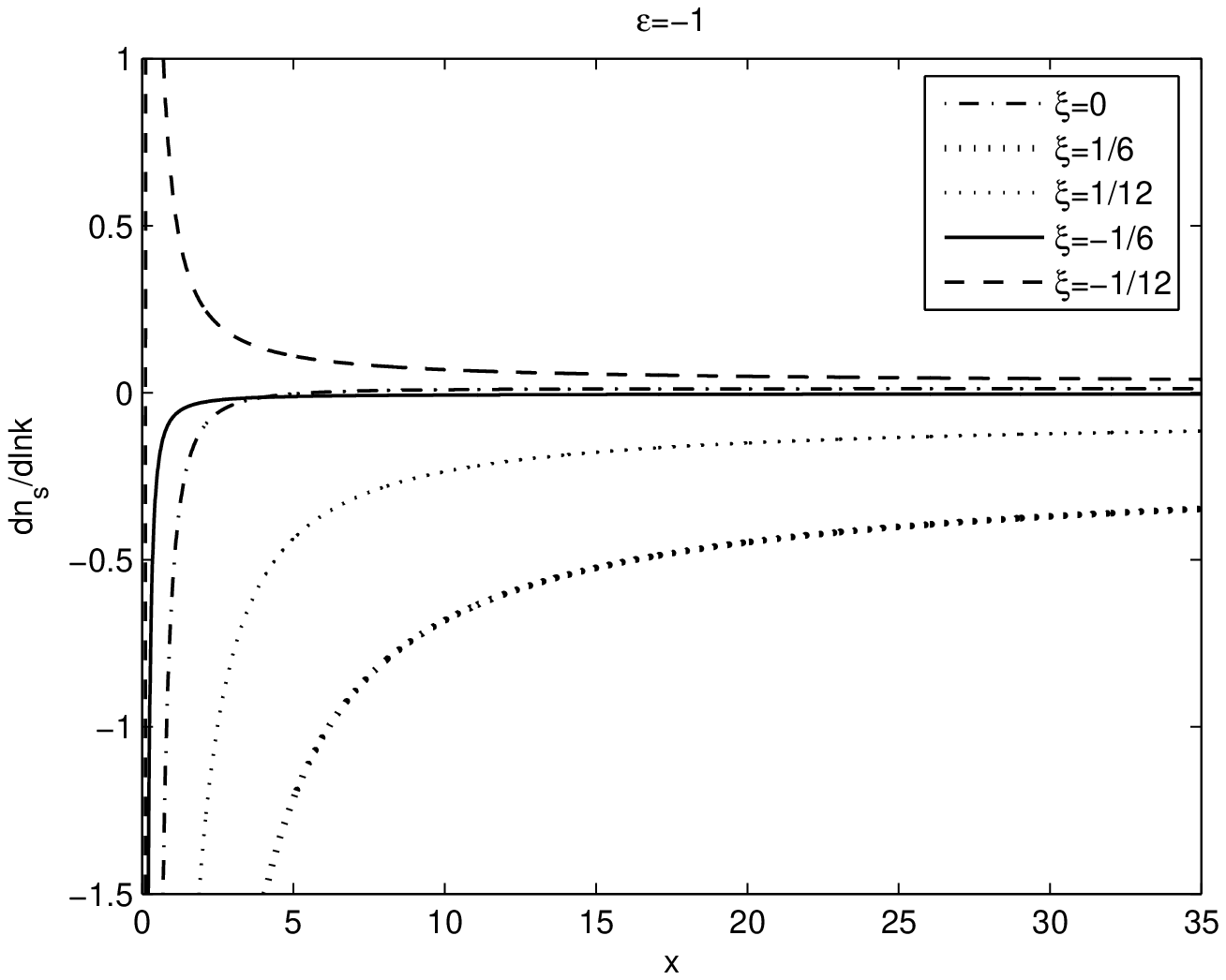}
\end{center}
\vspace{6 cm}
 \caption{\small { Running of the spectral index, $\frac{dn_s}{d\ln{k}}$
 for branch $\varepsilon=-1$}.}
 \label{fig:6}
\end{figure}

\begin{figure}[htp]
\begin{center}
\includegraphics{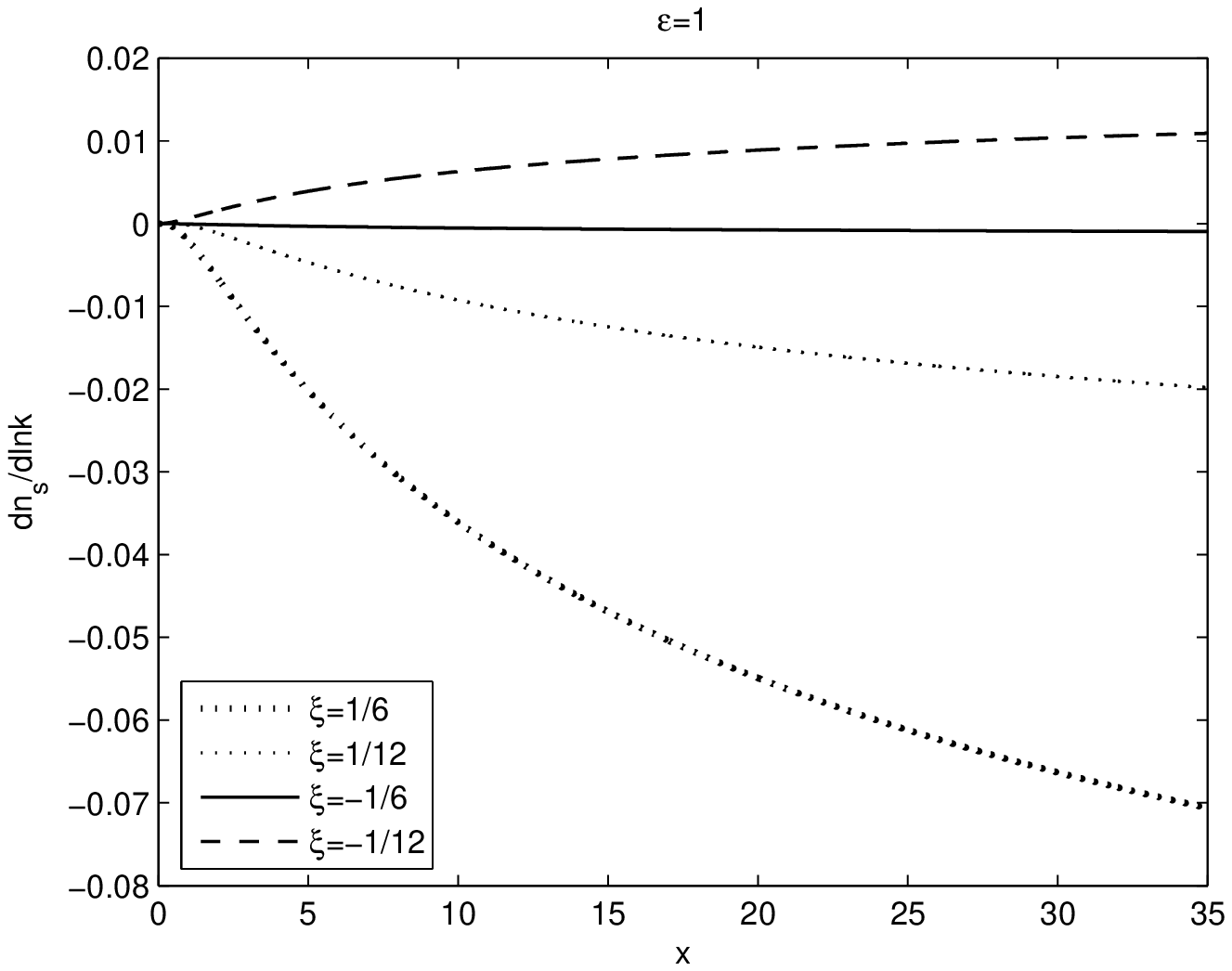}
\end{center}
\vspace{5 cm}
 \caption{\small { Running of the spectral index, $\frac{dn_s}{d\ln{k}}$
 for branch $\varepsilon=+1$}}
 \label{fig:7}
\end{figure}

\begin{figure}[htp]
\includegraphics{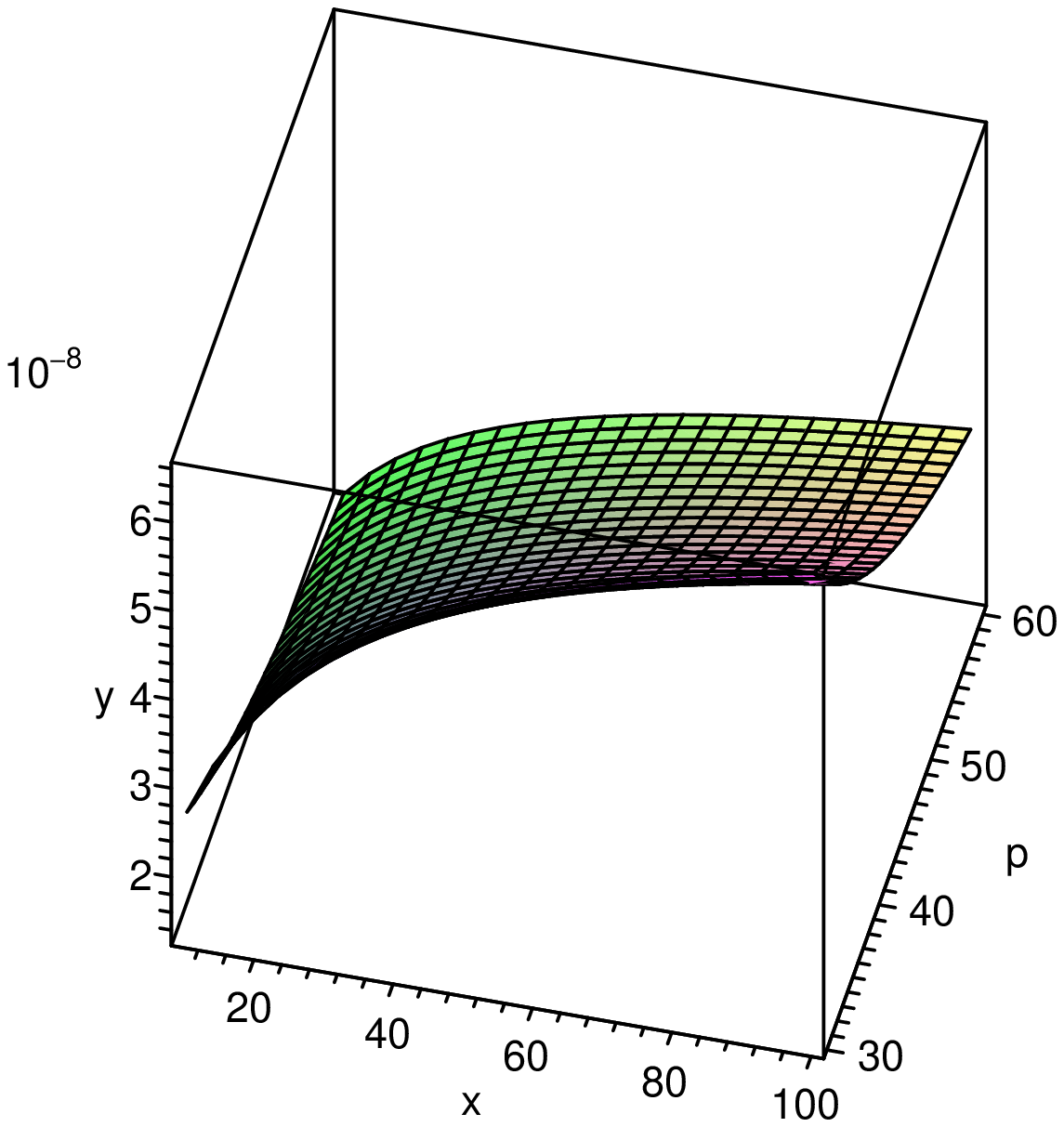} \vspace{7 cm}
 \caption{\small { Energy Scale of Non-Minimal Inflation with $\xi=\frac{1}{6}$}
 \hspace{11cm}}
 \label{fig:8}
\vspace{10cm}

\includegraphics{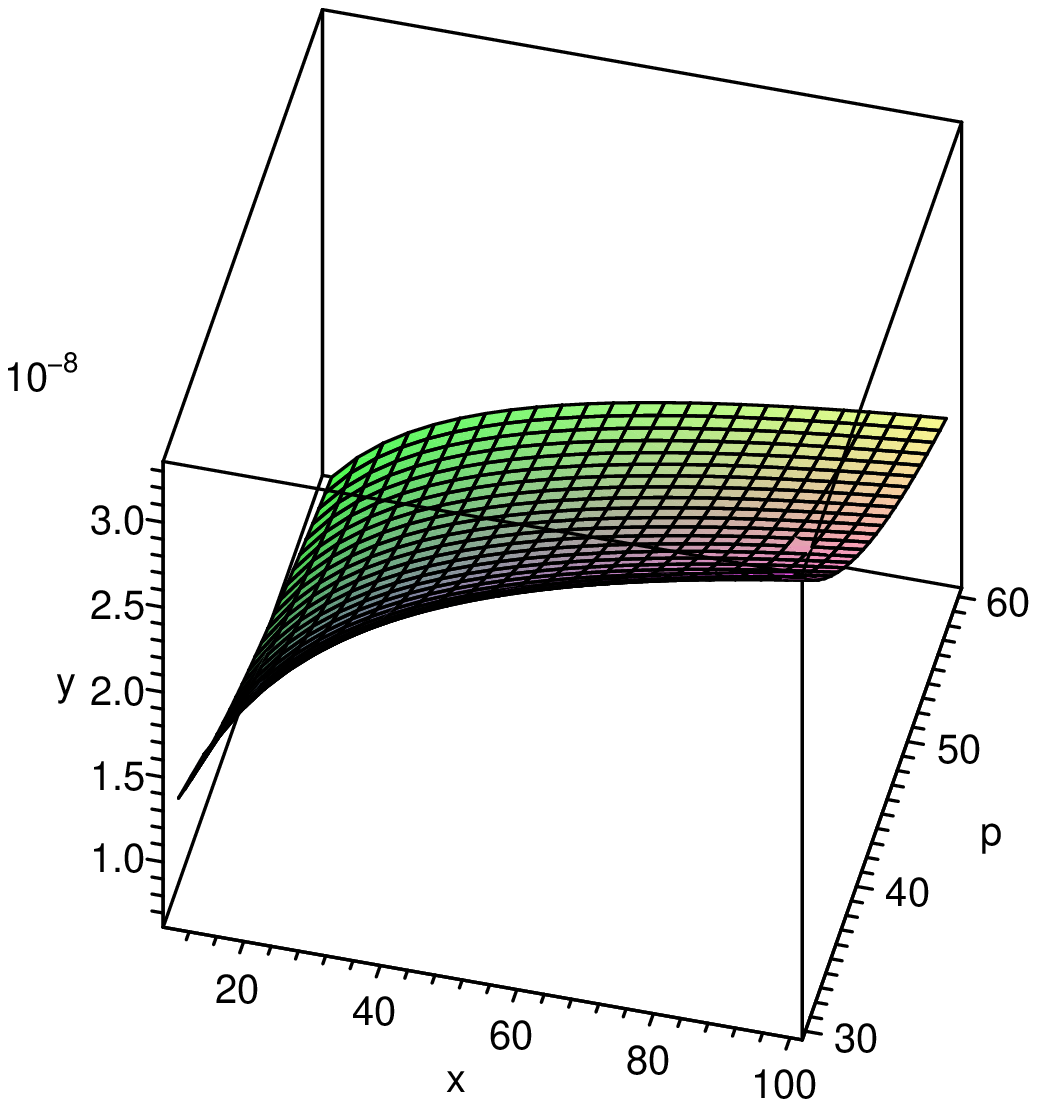}

\vspace{-11 cm}  \caption{\small {$\xi=\frac{1}{12}$}\hspace{-12cm}}
 \label{fig:9}
 \vspace{10cm}

\includegraphics{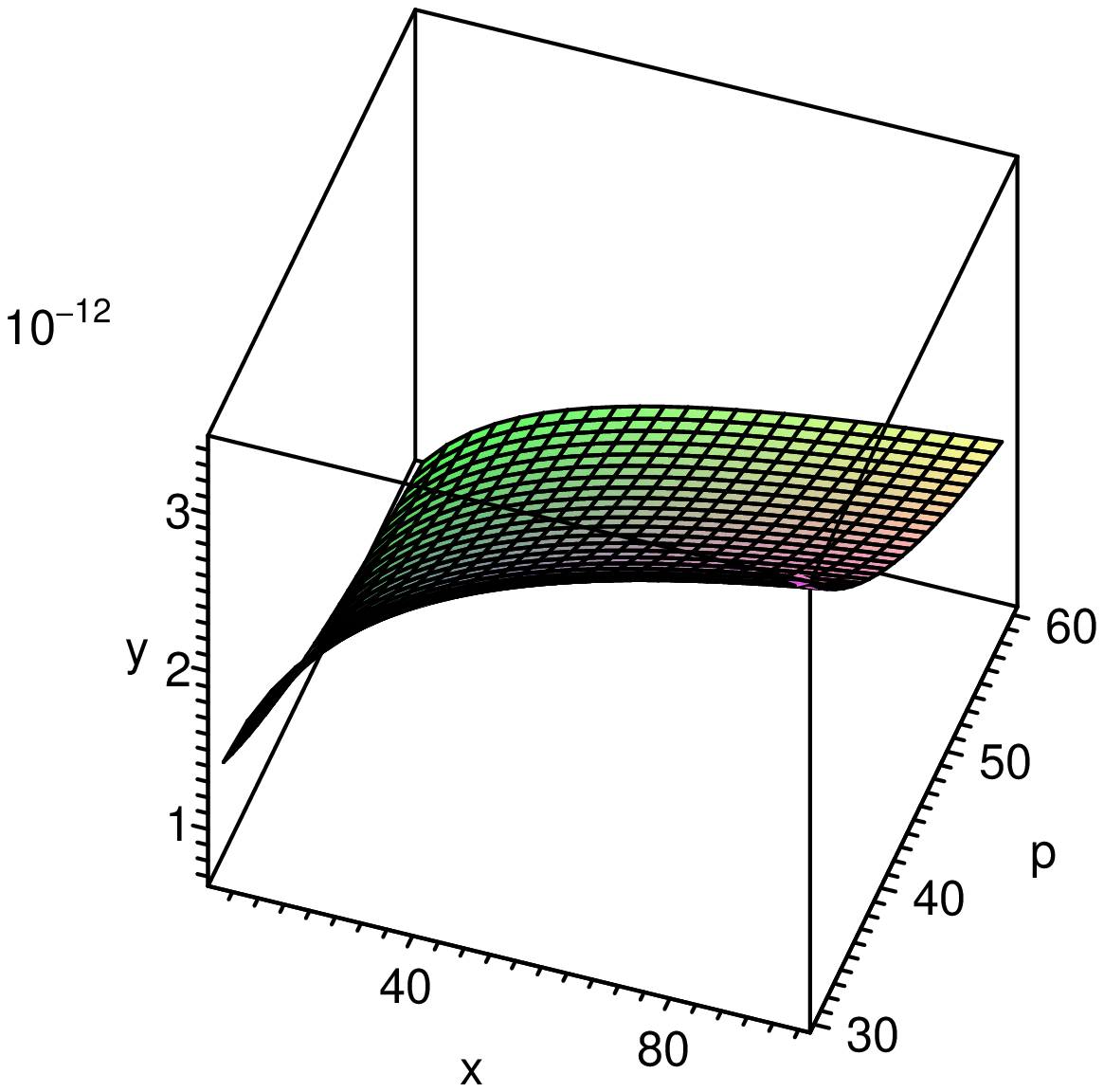}

\vspace{1 cm}
 \caption{\small { $\xi=-\frac{1}{6}$}\hspace{11cm}}
 \label{fig:10}

\begin{center}
\includegraphics{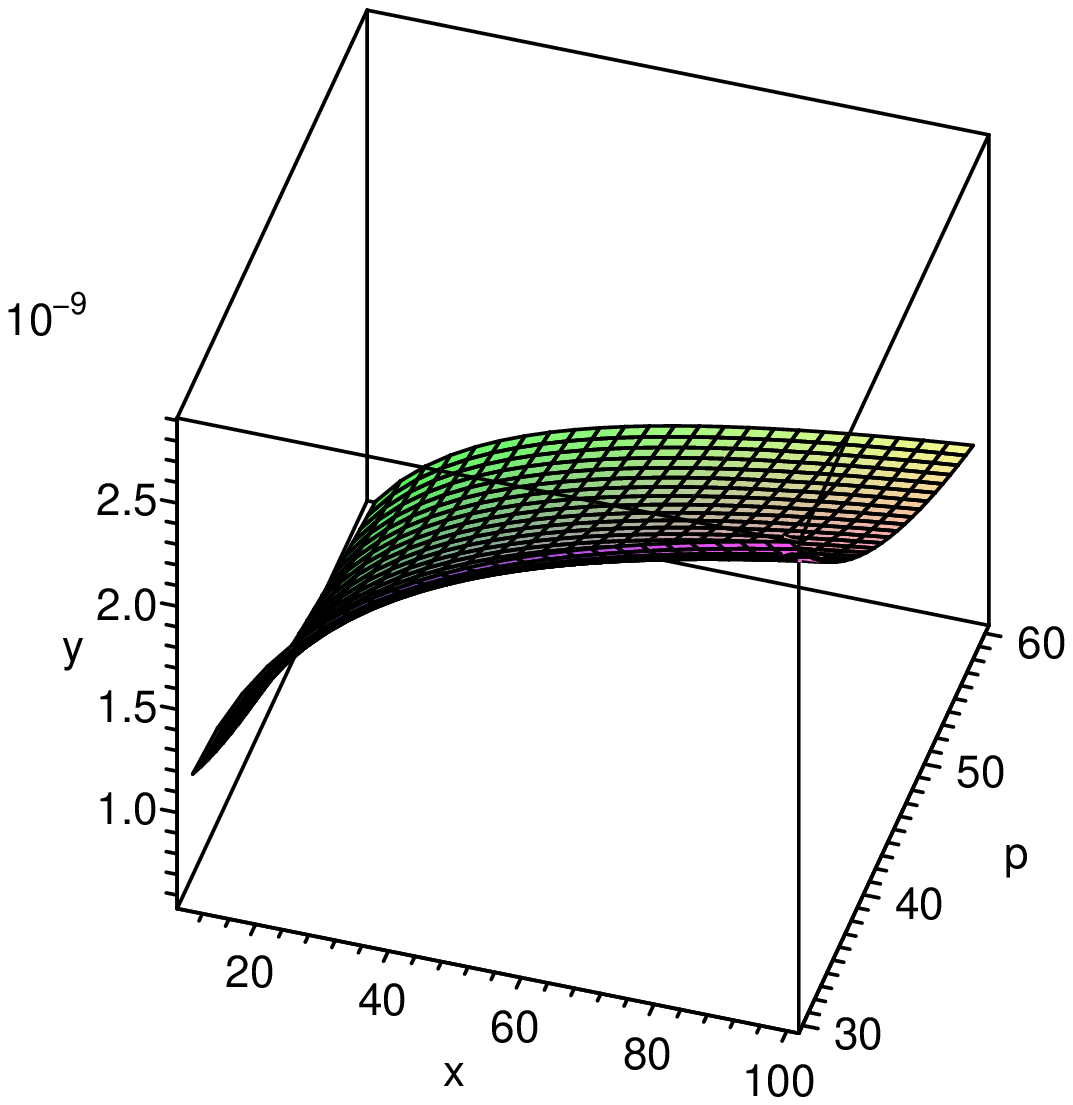}
\end{center}
\vspace{-2.5 cm}
 \caption{\small { $\xi=-\frac{1}{12}$}\hspace{-12cm}}
 \label{fig:11}
\end{figure}

\end{document}